\documentclass{IOS-Book-Article}

\usepackage{mathptmx}
\usepackage{soul}\setuldepth{article}
\usepackage{qtree} 

\usepackage{graphicx}

\def\hb{\hbox to 11.5 cm{}}

\usepackage[hidelinks]{hyperref} 
\hypersetup{
    colorlinks=true,
    linkcolor=blue,
    filecolor=blue,      
    urlcolor=blue,
    citecolor=blue,
    pdftitle={An ontological analysis of attack trees: Toward adequacy and semantic interoperability},
    pdfpagemode=FullScreen,
    }

\makeatletter
\renewcommand{\paragraph}{\@startsection{paragraph}{5}{0em}%
  {.0ex plus .2ex minus .1ex}%
  {-.5em}%
  {\bfseries}}
\makeatother

\begin{document}

\pagestyle{headings}
\def\thepage{}
\begin{frontmatter}              

\title{An ontological lens on attack trees: Toward adequacy and interoperability}

\markboth{}{April 2025\hb}





\author[A]{\fnms{Ítalo} \snm{Oliveira}\orcid{0000-0002-2384-3081}%
\thanks{Corresponding Author: Ítalo Oliveira, i.j.dasilvaoliveira@utwente.nl, \url{https://italojsoliveira.github.io}.}},
\author[B]{\fnms{Stefano Maria} \snm{Nicoletti}\orcid{0000-0001-5522-4798}},
\author[C,D]{\fnms{Gal} \snm{Engelberg}\orcid{0000-0001-9021-9740}},
\author[E]{\fnms{Mattia} \snm{Fumagalli}\orcid{0000-0003-3385-4769}},
\author[D]{\fnms{Dan} \snm{Klein}\orcid{0000-0002-8881-1902}}
and
\author[A]{\fnms{Giancarlo} \snm{Guizzardi}\orcid{0000-0002-3452-553X
}}

\runningauthor{Í. Oliveira et al.}
\address[A]{Semantics, Cybersecurity, \& Services (SCS), University of Twente, Enschede, The Netherlands}
\address[B]{Formal Methods and Tools (FMT), University of Twente, Enschede, The Netherlands}
\address[C]{University of Haifa, Haifa, Israel}
\address[D]{Accenture, The Center of Advanced AI, EMEA}
\address[E]{KRDB Research Centre on Knowledge and Data, Free University of Bozen-Bolzan, Bolzano, Italy}

\begin{abstract}
Attack Trees (AT) are a popular formalism for security analysis. They are meant to display an attacker's goal decomposed into attack steps needed to achieve it and compute certain \textit{security metrics} (e.g., attack cost, probability, and damage). ATs offer three important services: (a) conceptual modeling capabilities for representing security risk management scenarios, (b) a qualitative assessment to find root causes and minimal conditions of successful attacks, and (c) quantitative analyses via security metrics computation under formal semantics, such as minimal time and cost among all attacks. Still, the AT language presents limitations due to its lack of ontological foundations, thus compromising associated services. Via an ontological analysis grounded in the Common Ontology of Value and Risk (COVER)--- a reference core ontology based on the Unified Foundational Ontology (UFO)--- we investigate the ontological adequacy of AT and reveal four significant shortcomings: (1) \textit{ambiguous syntactical terms} that can be interpreted in various ways; (2) \textit{ontological deficit} concerning crucial domain-specific concepts; (3) \textit{lacking modeling guidance} to construct ATs decomposing a goal; (4) \textit{lack of semantic interoperability}, resulting in \textit{ad hoc} stand-alone tools. We also discuss existing incremental solutions and how our analysis paves the way for overcoming those issues through a broader approach to risk management modeling. 

\end{abstract}

\end{frontmatter}
\markboth{April 2025\hb}{April 2025\hb}


\section{Introduction}


ISO 31000~\cite{iso_31000} describes risk assessment as a process that includes risk \textit{identification}, \textit{analysis}, and \textit{evaluation}, which informs risk treatment decisions. Risk identification aims to identify and describe risks that might prevent an organization from achieving its objectives. It should consider factors such as tangible and intangible sources of risk, causes and events, vulnerabilities and capabilities, changes in the external and internal context, the nature and value of assets, consequences and their impact on objectives, and time-related factors, among others. Risk analysis considers uncertainties, risk sources, consequences, likelihood, events, scenarios, controls, and their effectiveness. Risk analysis techniques can be qualitative, quantitative, or both. It provides input to risk evaluation--- decisions on whether the risk needs to be treated, how, and the most appropriate risk treatment options. Numerous risk assessment techniques exist, and many are listed in ISO 31010~\cite{iso_31010}, including Failure Mode and Effects Analysis (FMEA), Fault Tree Analysis, Bow Tie Analysis, Markov Analysis, Monte Carlo Simulation, and Bayesian Statistics.

Attack Trees (AT) are a popular risk assessment technique designed to address security risks (attacks). They are formally defined as rooted directed acyclic graphs with typed nodes. They are meant to display an attacker's goal decomposed into attack steps needed to achieve it, and often analyzed in light of \textit{security metrics} (e.g., attack cost, probability, and damage). ATs offer three important services: (a) \textit{conceptual modeling} capabilities for representing security risk management scenarios; (b) \textit{qualitative analysis} capabilities to find root causes and minimal conditions of successful attacks; and (c) \textit{quantitative analysis} capabilities via security metrics computation under formal semantics (e.g., minimal time and cost among all attacks). However, the AT language has limitations due to its lack of ontological foundation, which affects the quality of AT models. Since AT qualitative and quantitative analyses are as good as the AT model, AT services are compromised by the lack of conceptual clarity. For example, calculating the minimal conditions of successful attacks is only useful if the representation of possible attacks is adequate.

Via an ontological analysis grounded in the \textit{Common Ontology of Value and Risk} (COVER)~\cite{sales2018common}--- a reference core ontology based on the \textit{Unified Foundational Ontology} (UFO)~\cite{guizzardi2022ufo_applied_ontology_journal}--- we investigate AT ontological adequacy and show four major sets of AT shortcomings: ($S_1$) \textit{a series of ambiguous syntactical terms} (semantically overloaded) that can be interpreted in various ways, such as the graph nodes that can be understood as goals, situations, events, event types, or even propositions; ($S_2$) \textit{ontological deficit} (incompleteness), meaning the AT language does not explicitly support crucial domain-specific concepts, including vulnerability, stakeholder, and asset; ($S_3$) \textit{lacking modeling guidance} to construct ATs decomposing a goal; ($S_4$) \textit{lack of semantic interoperability}, resulting in an \textit{ad hoc} stand-alone tool that can hardly be connected to data from multiple sources or other risk management techniques. We also discuss existing incremental solutions and how our analysis paves the way for overcoming those issues through a broader approach to risk management modeling.

Section \ref{attack_tree_description} describes the core ideas of static ATs, the target of our analysis. Section \ref{section_ontological_foundations} outlines the ontological foundations of risk based on COVER, which grounds our analysis. Section \ref{section_contribution_ontological_analysis} presents our contributions: an ontological analysis of static ATs revealing semantic issues $S_1-S_4$. Section \ref{section_discussion_solution} discusses our findings and moves toward a solution. Section \ref{section_related_work} discusses related works, particularly AT extensions attempting to overcome the lack of expressivity. Section \ref{section_conclusion} concludes the contribution.


\section{Attack trees: What are they? Where do they live?}
\label{attack_tree_description}



ATs are hierarchical diagrams that visually represent how a system can be attacked. Introduced in 1999 by Bruce Schneier \cite{schneier1999attack}, ATs are now popular both in industry and academia. They are referred to by many system engineering frameworks, e.g., \emph{UMLsec} \cite{Jur02} and \emph{SysMLsec} \cite{RA15}, are supported by industrial tools--- such as Isograph's \emph{AttackTree} and Amenaza's \emph{SecurITree}--- and academic tools, such as \emph{ADTool}. Collections of AT models are publicly accessible on online repositories\footnote{\url{https://dftbenchmarks.utwente.nl/at/ats.php}}. 

\begin{figure}[htb]
    \centering
    \includegraphics[width=0.35\textwidth]{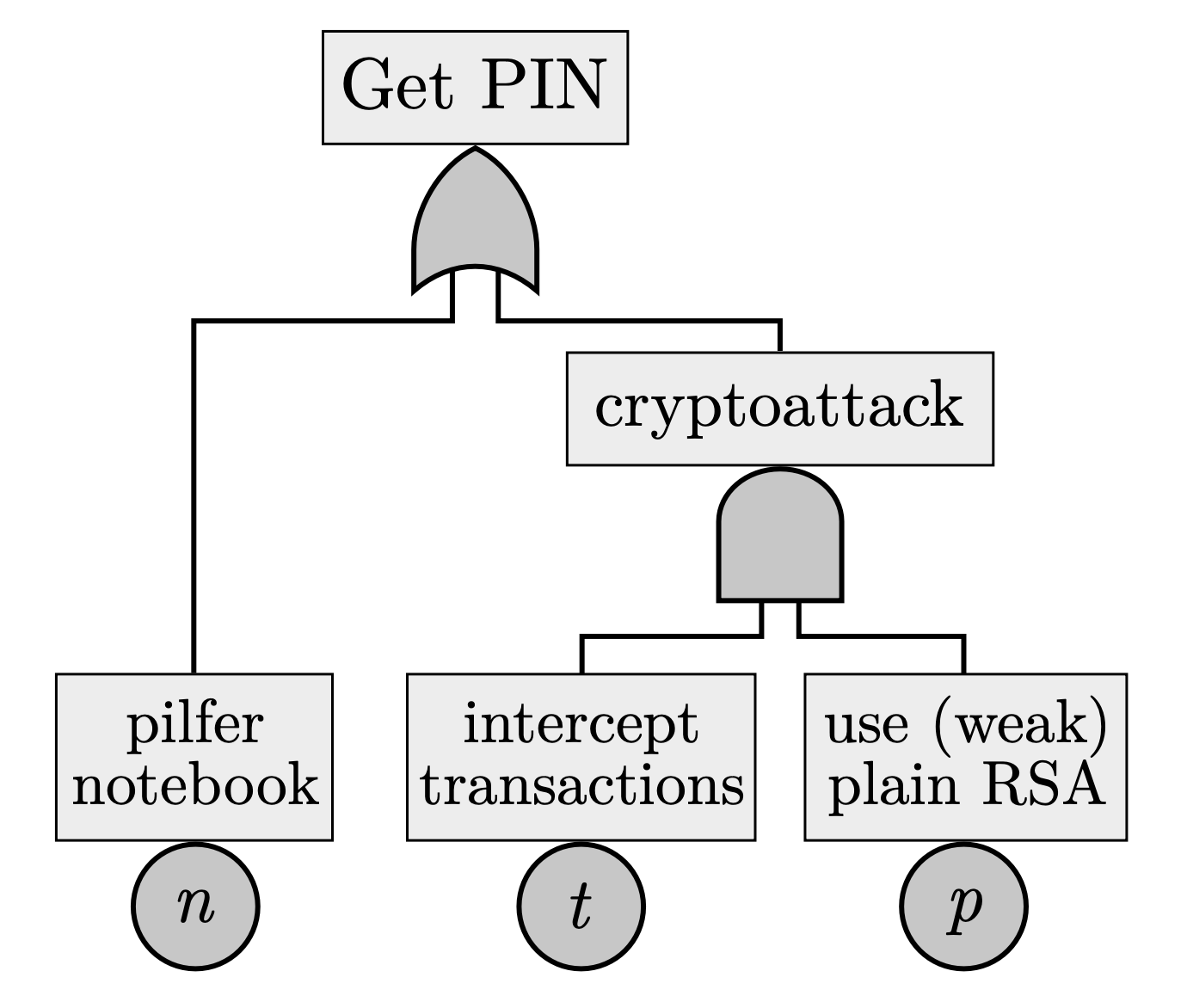}
    \caption{A simple AT representing attacks to obtain a digital PIN, from~\cite{lopuhaa2022efficient}.}
    \label{figure_simple_at}
\end{figure} 

Syntactically, ATs are rooted directed acyclic graphs (DAGs) with typed nodes \cite{lopuhaa2022efficient}. Figure \ref{figure_simple_at} depicts a simple AT from \cite{lopuhaa2022efficient}, representing attacks to obtain a digital PIN. According to Schneier \cite{schneier1999attack}, the \emph{root node} of the graph--- also known as \emph{top level event} (TLE) \cite{lopuhaa2022efficient}--- represents the overall \emph{goal} of the attacker (getting the PIN, in this example). This goal is then detailed into \emph{subgoals} via logic gates. Classic or \textit{static} ATs present only two types of gates, AND- and OR-gates. Our analysis focuses on this more widespread variant due to its popularity. We mention some AT extensions in Section \ref{section_related_work}. In Figure \ref{figure_simple_at}, to get the PIN the attacker needs to achieve at least one of two subgoals, as represented by the OR-gate that decorates \textit{Get PIN}: either they \textit{pilfer the notebook} of the victim, or they mount a \textit{cryptoattack} against the target. For this smaller attack to be successful, the attacker needs to both \textit{intercept transactions} where the PIN is used and \textit{weak RSA} needs to be used for the malicious agent to retrieve the information. This is represented by the logical AND-gate that decorates the \textit{cryptoattack} subgoal. Once the level of granularity of the model is deemed appropriate by the analyst, we arrive at the leaves of the graph, also known as \emph{basic attack steps} (BASes) \cite{schneier1999attack,lopuhaa2022efficient}. Whereas the TLE represents the attacker's goal, leaves represent--- in Schneier's words--- ``different ways of achieving that goal'' \cite{schneier1999attack}. No further details are provided on the nature of the relationship between nodes in the AT. Furthermore, the literature diverges about the interpretation of leaf nodes, sometimes understood as representing \textit{actions}, instead of subgoals. We will tackle these different interpretations in Section \ref{section_contribution_ontological_analysis}. 

Already at their initial formulation, ATs were intended to be studied both \emph{qualitatively} and \emph{quantitatively}. Qualitative analysis of ATs points out how different \emph{attacks} permit the achievement of the overall goal. Attacks are often represented by a (sub)set of the AT leaves \cite{lopuhaa2022efficient}, where each leaf element of the attack is assigned a value between 1--- meaning that the leaf subgoal/action is achieved/performed--- and 0, otherwise. A possible attack for the AT in Figure \ref{figure_simple_at} is represented by the set $\{n\}$ (we assume the convention that elements in the set are assigned value 1 and all the others are assigned 0). This means that an attacker is only trying to \textit{pilfer notebook} to achieve the overall goal, \textit{Get PIN}. This is a \textit{successful} attack given the structure of our example AT. This happens because $n$ is a direct child of \textit{Get PIN}, decorated with an OR-gate, for which only one of the children subgoals needs to be achieved for the attack to propagate. If we wanted to mount a different successful attack to get the PIN, we would need both \textit{intercept transactions} and \textit{use (weak) plain RSA}. Namely, $\{t, p\}$ is a successful attack because both the children of the \textit{cyberattack} AND-gate are needed for the attack to propagate and achieve \textit{Get PIN}. Both of these attacks are \textit{minimal} w.r.t. the TLE of the AT in Figure~\ref{figure_simple_at}: if any of the elements of these attacks is removed, they cease to be successful. This AT also has a non-minimal attack, namely $\{n,t,p\}$. In general, one could verify whether a given attack reaches a certain AT node by making use of its \textit{structure function}~\cite{lopuhaa2022efficient}, i.e., a function returning true iff an attack reaches a node, given the Boolean representation of the AT. For example, the Boolean representation of the AT in Figure \ref{figure_simple_at} would be $n \vee (t \wedge p)$.


Minimal attacks are also important for some semantical definitions of ATs: e.g., \cite{lopuhaa2022efficient} defines the semantics of a static AT as its set of minimal attacks. How one chooses to define the AT semantics--- i.e., what constitutes an attack--- is crucial for the definition and computation of \textit{security metrics}: such metrics formalize how well a system performs in terms of security and are essential when comparing alternatives or making trade-offs. Typical examples of metrics include the minimal time~\cite{KRS15,KSR+18}, minimal cost~\cite{BLP+06}, maximal probability \cite{JKM+15} of a successful attack, maximal damage, minimal skill, and maximal challenge~\cite{lopuhaa2022efficient}. While there exists literature that exclusively focuses on a single metric, the majority of research in the field of AT analysis adheres to the seminal work of Mauw and Oostdijk~\cite{mauw2006foundations}, who pioneered the formulation of AT metrics in terms of semirings--- an algebraic structure consisting of a set and two operators, corresponding to AND- and OR-gates. AT literature offers multiple, non-compatible definitions of both AT semantics and AT metrics \cite{mauw2006foundations, lopuhaa2022efficient}, and each comes with its own definition of how the structure of an AT and a semiring together define the AT's security value. Irrespective of these differences, none of the different presentations addresses the limitations of ATs associated with a lack of ontological foundations.


\section{Ontological foundations of risk}
\label{section_ontological_foundations}


The Common Ontology of Values and Risk (COVER)~\cite{sales2018common} is a core ontology based on the Unified Foundational Ontology (UFO)~\cite{guizzardi2022ufo_applied_ontology_journal}. UFO differentiates the categories of \textsc{Types} (Universals) and \textsc{Individuals}, which are mutually disjoint. A car, as a type of object,t may have subtypes, such as a Honda car and a Fiat car. A particular car is an individual instantiating a car as a first-order type. Individuals can be \textsc{Abstract Individuals} (specific numbers, sets, propositions) or \textsc{Concrete Individuals} (existing in space-time). \textsc{Concrete Individuals} are disjointly classified into \textsc{Events}, \textsc{Situations} (individual state of affairs), \textsc{Objects}, and \textsc{Aspects} (or \textsc{Moments}). The latter include \textsc{Dispositions} and \textsc{Qualities}. Examples of \textsc{Dispositions} are Mary's ability to speak English, John's vulnerability to extortion, and Anna's liability for the damage she caused with her car. Examples of \textsc{Qualities} are Mary's height, and John's eye color. \textsc{Qualities} can inhere in other \textsc{Aspects}, e.g., the proficiency of John's hacking skills~\cite{guizzardi2022ufo_applied_ontology_journal}.

\textsc{Events} represent changes from one \textsc{Situation} to another. \textsc{Events} are manifestations of \textit{interacting} \textsc{Dispositions} in a given world setting. The following pattern emerges from this: certain \textsc{Situations} activate certain \textsc{Dispositions}; these are manifested by \textsc{Events} wherein \textsc{Objects} (the bearers of those \textsc{Dispositions}) participate, bringing about new \textsc{Situations}. We say that a \textsc{Situation} \textit{triggers} an \textsc{Event} if it activates the associated \textsc{Dispositions}. Examples of this pattern are: hackers (\textsc{Objects}) who acquired their skills (\textsc{Dispositions}) through training (\textsc{Events}); cyber attacks (\textsc{Events}) are manifestations of those skills exercised in particular scenarios, exploiting the target's vulnerabilities. Training and attacks modify the state of affairs, for example, by creating, eliminating, or altering \textsc{Objects} and \textsc{Aspects}. Unlike these, \textsc{Events} cannot change while keeping their identity: thus, they can only exist in the past~\cite{guizzardi2013towards,botti2019representing_ufo_b_dl,prevention_rcis21}.

According to COVER~\cite{sales2018common}, (1) \textit{risk is relative}: an event perceived as a \textsc{Threat Event} by one agent (\textsc{Risk Subject}) might be perceived as an \textsc{Opportunity Event} by another agent. Cyberattacks may cause \textsc{Loss Events} for a company but \textsc{Gain Events} for hackers. (2) \textit{The relativity of risks emerges from their negative impact on agents'} \textsc{Intentions} (goals). An actual loss means there is at least one agent whose goals have been hurt by events, which, therefore, are perceived as losses. \textsc{Threat Events} can cause \textsc{Loss Events} (both subtypes of \textsc{Risk Events}). (3) \textit{Risk is experiential}. Although we informally speak of \textsc{Objects at Risk} (e.g., a sensitive database that can be compromised) and \textsc{Threat Objects} (e.g., a hacker), the risk is derived from \textsc{Risk Assessment} of conceivable \textsc{Risk Experiences} composed of \textsc{Risk Events} wherein those objects participate. For example, to identify and assess the risks an information system is exposed to, we must describe: (i) which objectives or goals depend on this information system's functionalities (e.g., making a medical appointment); (ii) what can happen to this information system that could jeopardize its capability to serve those goals; (iii) which events could yield such a result (e.g., DDoS attacks, phishing, SQL injection, ransomware, etc.) - the risk that information system is exposed to is the aggregation of these risks. (4) \textit{Risk is contextual}: Risk depends on the intrinsic properties of objects (capabilities, vulnerabilities, intentions) but also depends on external contextual conditions (\textsc{Threatening Situations}), such as driving a car in a snowstorm or working without safety equipment. These \textsc{Situations} can activate certain \textsc{Threat Capabilities} of \textsc{Threat Objects} (v.g., a slippery road) and \textsc{Vulnerabilities} of \textsc{Objects at Risk} (e.g., a driver) (along with other required \textsc{Dispositions}), giving rise to \textsc{Risk Events} (e.g., car accidents followed by injuries). (5) \textit{The existence of risks depends on uncertainty about events and their outcomes}. This is why risk is often interpreted as the product of the \textit{likelihood} of certain \textsc{Risk Events} and the \textit{impact} of corresponding losses. Likelihood values can only be assigned to event types\footnote{COVER has two concepts of likelihood assignments: (a) \textsc{Triggering Likelihood} and (b) \textsc{Causal Likelihood}. The former defines the probability of a \textsc{Disposition}'s manifestation in certain \textsc{Situations}. The latter concerns event succession in a causal chain.}, not individual events, because these have already happened. 

While risks concern possible events in the future, incidents are realized instances of risks, i.e., occurred events. Consequently, risk analysis and incident analysis are analogous activities. The former analyzes potential occurrences, while the latter examines historical events. However, both analyses aim to inform risk treatment decisions, i.e., the actions to be taken regarding risks.


\section{Ontological analysis of attack trees}
\label{section_contribution_ontological_analysis}


The use of ontologies to evaluate and (re)design conceptual modeling languages has a long and illustrious tradition (~\cite{WandWeber1989,guizzardi2007ontology}). In this approach, a conceptual representation of the modeling constructs available in a language (sometimes called the \textit{ontological metamodel of the language}) is systematically confronted with an ontology taken as a reference model. The rationale is that the language's metamodel should be isomorphic to that reference ontology, i.e., the interpretation from the modeling primitives to the ontology concepts should be bijective. Failing to comply with this requirement yields ontological limitations affecting the language's expressivity and pragmatic efficiency \cite{guizzardi2007ontology}. These are: (a) \textit{ontological incompleteness} (\textit{construct deficit}), which is the lack of a grammatical construct for an existing ontological concept--- the language cannot easily or explicitly represent this concept; (b) \textit{construct overload}, which means that one construct represents more than one ontological concept--- introducing ambiguity into the models; (c) \textit{construct redundancy}, which happens when more than one construct represents the same concept--- more complexity without any expressivity gain; (d) \textit{construct excess}, which means a grammatical construct does not map to any concept--- lack of control over the (possibly multiple) interpretations assigned to that construct.

Taking UFO and COVER as our reference ontologies for an ontological analysis, we show that the formal definition of the static AT language exhibits examples of these limitations. Particularly, construct deficits and overloads affect AT language in multiple ways. Although ATs are designed to support risk assessment, they lack domain-specific language elements, and the interpretation of their constructs in terms of domain concepts is merely implicitly assumed. As a result, there is no real control over how the symbols in ATs are to be interpreted. This impacts AT modeling capability and hinders modeling guidance and interoperability.

\begin{figure}[ht]
      \includegraphics[width=0.55\textwidth]{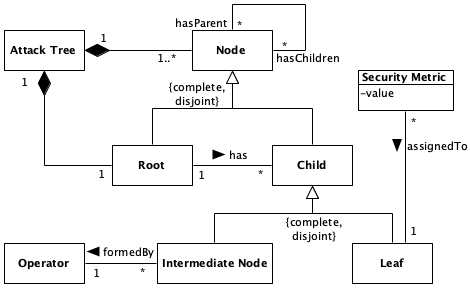}
      \centering
     \caption{A representation of the AT symbols in the UML Class Diagram.}
      \label{figure_AT_UML}
\end{figure}

Aligned with the AT metamodel defined by Kumar et al.~\cite{KSR+18}, we propose a UML Class diagram representation of the AT graph language, depicted in Figure \ref{figure_AT_UML}. This will help us understand the different roles of nodes and identify and visualize the ontological issues. Accordingly, one AT instance can be composed of several nodes, one of which is the root, whereas others are children connected (directly or indirectly to this root. A child node can be either an intermediate node or a leaf. Intermediate nodes are formed by logic gates. Security metrics are assigned to the leaves. An instance of a security metric is, for example, the value \$10.0 assigned to the ``break Wi-Fi password'' leaf node.

\subsection{Semantic overload}
\label{subsection_semantic_overload}


The following analogy clarifies general semantic overload issues of the AT formalism: it is like an algebraic statement such as $ x = y \cdot z $. Simultaneously, the AT language is only used to address domain-specific concerns, such as in the physical equation $ F = m \cdot a $. In the latter case, we know we are dealing with the resulting force yielded by the mass of an object and its acceleration, considering velocity and time. The underlying ontology is clearer, although not formalized. In the former case, we simply have an abstract equation, which is flexible enough to capture the latter case and many others we may not be interested in. Because of that, technically, nothing prevents ATs from being used as, for example, a goal-modeling language, such as i* framework~\cite{yu2011social}, unrelated to security risk management. This is why implicit domain-specific knowledge is necessary to make sense of minimal attacks and security metrics, which only matter in the security risk management context. E.g., in computing the metric for the maximal probability of successful attacks, the common interpretation of encountered ORs is given by the max operator. This is because practitioners assume that, given the choice between \textit{this or that} steps, the attacker will choose the one that maximizes their chances of success. However, this necessary domain knowledge remains implicit.

\paragraph{Node ambiguity.} The AT literature consistently identifies the root node as the attacker's final goal. However, this is not the case for the child nodes. They are referred to as ``steps'', ``attack steps'', ``subgoals'', ``undesired events'' (from the \textsc{Risk Subject}'s perspective, which is another implicit assumption), ``ways of achieving the goal'', ``attacker actions'', to name a few~\cite{schneier1999attack, mauw2006foundations, lopuhaa2022efficient,el2012modeling,KRS15}. It is worth noting that, from the attacker's perspective, attack steps are actually desired (\textsc{Opportunity} or \textsc{Gain Events}). The concept of a node in the AT language can denote any sort of entity. This generality is a double-edged sword because nodes usefully accommodate generic algorithms to compute different metrics. Cost, time, skills, challenge, damage, and probability can all be assigned to a leaf node. The same algorithm works for each metric. However, this flexibility has downsides as nodes can be interpreted as \textsc{Intentions} (\textsc{Intrinsic Aspects}), goals (propositions), \textsc{Objects}, \textsc{Situations}, \textsc{Events} (attacker's actions), or \textsc{Event Types}. This ontological confusion indeed happens in the same AT and across different ATs. Consider the following examples of nodes in an AT made by Bouchti and Haqiq~\cite{el2012modeling} having ``SCADA Compromised''\footnote{SCADA is an acronym for Supervisory Control and Data Acquisition, which is a control system architecture.} as the root node: (a) ``power loads not provided'', which looks like an occurred \textsc{Event}; however, it is not clear what exactly happened because the node description just says what did \textit{not} happen; (b) ``Incorrect monitoring'' and ``Incorrect estimates to customers'', which may be \textsc{Event Types} referring to possible sets of \textsc{Events}; (c) ``Incomplete sensors'' and ``Unavailable Network'', which may refer to \textsc{Situations} wherein sensors are incomplete and the network is unavailable; (d) ``Database'' and ``Workstation'', which seem to denote \textsc{Objects} somehow related to `Incorrect estimates to customers'' (their parent node). 

Kumar et al.~\cite{KRS15} draw an AT to model the forestalling of software (root). Intermediate nodes include ``Bribing'', ``Network attack'', and ``Physical robbery''. Some leaves are ``Hire robber having knowledge in computer security'', ``Bug in a computer system'', and ``Robber breaks into system''. These examples reveal that even the lexical category of nodes varies since they can use single verbs, nouns, and entire phrases. \textit{The description of the node does not have an internal structure or a meaning denoting a particular entity}. It is merely a label or annotation, such as \textit{n}, \textit{t}, and \textit{p} in Figure \ref{figure_simple_at}; these are shorthands for the descriptions, respectively, ``pilfer notebook'', ``intercept transactions'', and ``use (weak) plain RSA''. 

Because ATs have a single root node as an ultimate goal, they implicitly commit to the proposition that either only one attacker (\textsc{Threat Object}) is pursuing this goal or multiple attackers share the same goal. We can only make this distinction in ATs by describing the nodes, hence carrying no machine-readable formal meaning. This means that the node description itself is highly overloaded because, without good descriptions, we do not know what we are talking about. Consequently, \textit{the utility of calculating minimal conditions of successful attacks and security metrics depends entirely on naming nodes well enough}.

\paragraph{Edge and Gate ambiguity.} The ontological semantics of the edges and gates (respectively, the \textit{has} and \textit{formedBy} associations of Figure \ref{figure_AT_UML}) are also overloaded. Take the example of Figure \ref{figure_simple_at}. ``Get PIN'' is the name of the formula $n \vee (t \wedge p)$, which can be better understood as: ``pilfer notebook'' OR (``intercept transactions'' AND ``use (weak) plain RSA''). A ``cryptoattack'' is the name of the subformula (``intercept transactions'' AND ``use (weak) plain RSA''). What is the relationship between ``intercept transactions'' and ``use (weak) plain RSA''? And what is the relationship between them and ``cryptoattack''? Some possibilities are: (a) a cryptoattack is an event composed of at least one transaction interception and one plain use of weak RSA; therefore, the AND-gate denotes a \textit{mereological relation between events}; the compositionality of the propositional formula suggests this interpretation; (b) a cryptoattack is an event caused by at least one transaction interception event and one plain use of weak RSA event, in conjunction; so, the AND-gate would mean a \textit{strong historical dependence relation} (specific causation)\footnote{Our analysis only considers static AT that does not account for temporal aspects or sequentiality. Dynamic ATs introducing SAND-operator overload the language even more.}; (c) this causation may be understood as a type-to-type relation referring to \textit{regular causal patterns} among possible events, instead of a specific dependence; (d) the relationship may be understood as the parthood description of an attacker's \textsc{Intention}: an attacker is committed to executing a cryptoattack, which is an \textsc{Intrinsic Aspect} composed of the parts ``intercept transactions'' and ``use (weak) plain RSA''; so and AND-gate would denote \textsc{Aspect} (\textsc{Endurant}) \textit{parthood}; this is also justified by  the compositionality of the propositional formula; (e) the AND-gate can be interpreted more literally in the sense that it refers to a truth function applied to a \textit{proposition} ``cryptoattack'' (composed of ``intercept transactions'' $ \wedge $ ``use (weak) plain RSA''); to make sense of this in ATs, we have to say this proposition is the goal of an attacker's \textsc{Intention}; then, an entire AT would be understood as a nothing more than a single proposition, missing domain-specific concepts described in Section \ref{section_ontological_foundations}; (f) finally, the relationship can denote the impact of events on goals (\textsc{Intentions}) so ``intercept transactions'' and ``use (weak) plain RSA'' could be necessary events to achieve the ``cryptoattack'' goal; an AND-gate would represent necessary contributions to reach a goal (or \textsc{Intention}); this interpretation either leaves open which nodes are goals (or \textsc{Intentions}) and which are events, or collapses events and goals/\textsc{Intentions}.

\paragraph{Security metric meaning.} Given the node ambiguity, security metric assignments face a similar semantic overload because they are \textsc{Qualities} assigned to different types of entities (collapsed into the leaf notion). For example, depending on the exact interpretation, all security metrics, such as probability and costs, could be assigned to types of \textsc{Events} (sets of events) or types of \textsc{Situations} (sets of situations). However, a probability cannot be assigned to individual events. Moreover, although certain types of events may have a minimal duration, only individual events have an exact duration in space and time. Similarly, what does it mean to assign a cost or skill level to a goal? The minimal skill required to execute an AT step may refer to a \textsc{Quality} (skill level) of a \textsc{Disposition} (skill), which inheres in a \textsc{Threat Object} (attacker).


\subsection{Ontological incompleteness}
\label{subsection_semantic_incompleteness}


The AT formalism does not have the concepts necessary to understand the risk domain while implicitly assuming some of them--- without which ATs simply do not make sense beyond symbolic computation. In what follows, we show exactly what is missing and the consequences of that.

\paragraph{Security risk scenario illustration and construct deficit.} Consider the following security risk scenario: a burglar (intentional \textsc{Threat Object}) is part of a criminal organization (intentional \textsc{Threat Object}), whose business involves assaulting houses (\textsc{Object at Risk} or an enabler) to steal (\textsc{Risk Event}) electronic equipment (\textsc{Object at Risk}) and resell them in the black market. The criminal organization selects its most skilled members (\textsc{Threat Capabilities} with \textsc{Quality} value ``high'') to assault the best-protected (\textsc{Vulnerabilities} with \textsc{Quality} value ``low'') houses in the region. The burglar has to identify and exploit physical and social vulnerabilities related to the house, such as unlocked doors and the residents' personalities. Successful attacks (\textsc{Threat Events} that caused \textsc{Loss Events}, from the victim's perspective) target specific assets inside the house and generate benefits satisfying the organization's goals. On the other hand, those represent damage to the \textsc{Risk Subject}'s \textsc{Intention}--- the commitment to protecting their assets. Certain \textsc{Situations} favor successful assaults, such as when nobody is in the house (\textsc{Threatening Situation}), which usually happens on a given weekday (likelihood assignments). The criminal organization and the burglar have different \textsc{Intentions} (each \textsc{Aspect} inheres in a single individual) but may share a propositional content (accomplishing successful assaults).

The theory of risk, described in Section \ref{section_ontological_foundations}, explains \textit{why} a successful attack occurs, \textit{who} are the ones affected and \textit{how} they are affected, which objects participate in attacks, the \textit{role} and \textit{features} of capabilities and vulnerabilities in certain situations. The AT language cannot distinguish any of this because it lacks explicit domain-specific concepts and an underlying theory of risk. As a result, a complete AT without contextualization and documentation can hardly be understood. For example, here is a minimal successful attack of AT with 24 total nodes (leaves included): ``SCADA Compromised'' (root), ``Incorrect estimates to customers'' (child), ``Database'' (leaf)~\cite{el2012modeling}. From the attacker's perspective, which is how AT is meant to be read, to achieve ``SCADA Compromised'', the attacker should \textit{verb\_placeholder} ``Incorrect estimates to customers'', which requires that the attacker \textit{verb\_placeholder} ``Database''. Without further instructions and details, even constructing a grammatically correct English description out of this is a difficult task. The simple example of Figure \ref{figure_simple_at} is not better: why is ``Get PIN'' relevant for an attacker? What are the consequences of the realization of ``Get PIN''? Are we considering a single or multiple attackers? Who is interested in protecting this PIN, that is, who is the \textsc{Risk Subject} (stakeholders)? What conditions and vulnerabilities favor an attack? These are just a few unanswered questions by the AT, although they are crucial parts of the risk assessment process, according to ISO 31000~\cite{iso_31000}.


\subsection{Limited modeling guidance}
\label{subsection_limited_modeling_guidance}



The semantic overload and ontological incompleteness of the AT language have the consequence of posing a \textit{heavy workload on AT users} (security analysts). As we have seen, the node description (e.g., ``Get PIN'')--- a conceptual modeling task--- is the most important activity in making AT diagrams. We can only know what we are calculating if we describe security risk scenarios. We can only calculate useful things (minimal attacks and security metrics) if we describe security risk scenarios \textit{rightly}. Nevertheless, the AT language offers little to no conceptual modeling guidance, increasing the chances of creating imprecise models. AT users have to (a) come up with the attacker's final goal (root node) and start to (b) branch it into intermediate nodes until (c) they reach ``basic attack steps''. These are nodes that cannot be further decomposed. How to do this and whether the model is correct are completely up to the AT users to figure out. To workaround this problem, it is possible to map AT nodes to known security risk taxonomies, such as techniques of the MITRE ATT\&CK and D3FEND~\cite{husseis2023enhancing}. However, this does not fix the semantic overload and incompleteness of the AT language and relies on the ontological adequacy of the taxonomy--- and it has been demonstrated that MITRE ATT\&CK and D3FEND exhibit deficiencies in their respective ontologies~\cite{oliveiraboosting}.

Modeling guidance depends on \textit{modeling patterns} reflecting a domain ontology~\cite{falbo2016ontology}. These patterns imply that language elements cannot appear in isolation but only following grammatical constraints designed according to a domain ontology. For example, according to the theory of risk of Section \ref{section_ontological_foundations}, an instance of an \textsc{Intention} (say, the attacker's commitment to getting a PIN) implies the existence of its bearer (the attacker); an attacker is classified as such only because of its capabilities and participation in events that can hurt someone else's objectives--- hence, implying the existence of stakeholders, assets, goals, and respective events. This is the modeling guidance missing in ATs and that is present in other domain-specific modeling languages, such as the ArchiMate's Risk and Security Overlay~\cite{oliveira2024ontology}.


\subsection{Semantic interoperability issues}



The ontological issues we have seen so far are detrimental to the conceptual modeling services offered by ATs and, consequently, to their qualitative and quantitative risk assessment. However, from a broader perspective, we also observe a major manifold impact on AT interoperability, turning ATs into an \textit{ad hoc} stand-alone tool.

\paragraph{Data interoperability.} The first interoperability problem regards building ATs based on existing public and private datasets and knowledge bases. The lack of conceptual clarification blocks a systematic approach to this task because we have to decide how to map various datapoint types to the AT language in each case. For example, given data obtained from different systems and organizations, e.g., internal security log records, MITRE ATT\&CK, and CAPEC attack patterns taxonomy, how should we generate adequate ATs from that? Instead, suppose, for the sake of argument, we settle the ontological nature of AT nodes as types of \textsc{Threat Events}. Then, this is exactly the type of entity we have to look up in the data sources. Good domain ontologies provide the conceptual clarity needed for this interoperability.

\paragraph{Risk assessment technique interoperability.} The second interoperability issue concerns the integration of different risk assessment techniques, as listed in ISO 31010~\cite{iso_31010}. Can we construct ATs by retrieving information from a Bayesian model, a fault tree, an FMEA analysis, a risk matrix, or other risk assessment techniques? What, if any, are the relationships between these approaches? These are not original concerns of ATs but answering these questions can greatly improve security risk analyses. Such technique integration would provide risk assessment analyses from different perspectives.

\paragraph{Human communication.} A third issue related to interoperability involves human communication. Due to those ontological limitations, we hypothesize that \textit{two persons making an AT separately for describing a given security risk scenario will create completely different ATs}. As shown in a previous example~\cite{el2012modeling}, sharing an AT model with someone without providing further instructions, which depends on the AT creator's communication skills, is not an effective means of communication. Besides the lack of understanding, there is the risk of false agreement. In fact, there is a problem with the very AT identity criteria: when are two ATs the same? Because the node description is not part of the AT language, two ATs can look exactly the same, e.g., having the form $n \vee (t \wedge p)$, although their \textit{intended meanings} differ drastically. As the AT identity criteria are purely formal, the same person may be puzzled about the meaning of the same AT over time.

%


\section{Toward interoperable and ontologically adequate operation of attack trees}
\label{section_discussion_solution}


The AT formalism can be seen as a Domain-Specific Language (DSL) for security risk management that fails to be isomorphic to the domain ontology. ATs exemplify symbolic models that are not explanatory due to those ontological problems. As argued by Guizzardi and Guarino~\cite{GUIZZARDI202410232}, the mere fact that an artifact is a symbolic description does not make it explanatory. A symbolic structure that is also explanatory must reveal the truthmakers of the description, i.e.,  \textit{why} (under which conditions in the world) that is the case, not just that \textit{that} is the case.

\paragraph{Incremental solutions.} Can we improve AT capabilities, making them more interoperable, ontologically adequate, and explanatory models? Researchers have tried to increase AT expressivity by adding certain elements to the language--- for example, a notion of defense or countermeasure~\cite{kordy2014attack} and sequentiality~\cite{lopuhaa2022efficient} (see Section \ref{section_related_work}). Based on our analysis, we could propose, for instance, the addition of more than one root node and assigning each of them to an attacker element to express explicitly attackers and their goals. This would address one of the highlighted issues in Subsection \ref{subsection_semantic_incompleteness}, relative to construct deficits. Although functional in principle, this would be an incremental solution that creates a new formalism for which current AT algorithms would need to be adjusted. Most importantly, without the support of a proper domain ontology, these language increments tend to increase the semantic overload (ambiguity). This bottom-up approach of fixing issues by extending the expressiveness of an AT model must be accompanied by a top-down approach enabling the grounding of (extended) models into a sound ontology underneath.

\paragraph{Towards a general framework.} We propose such a general top-down solution that includes and goes beyond ATs. Our proposal is a systematic approach to risk management combining Conceptual Modeling and Applied Ontology, Formal Methods, and Semantic Web technologies, following~\cite{oliveira2025toward}. Here we focus on how it can address AT shortcomings. Essentially, there is no way around an ontological analysis of the domain of risk and security. References ontologies such as COVER and the \textit{Reference Ontology for Security Engineering} (ROSE)~\cite{oliveira2022security} are necessary to support the creation of any explanatory artifacts for risk management. The concept of risk propagation--- widely used in several risk assessment techniques--- requires a similar analysis as it is highly overloaded~\cite{fumagalli2023semantics}. \textit{ATs should be conceptually defined in terms of these well-founded risk and security models}. This means that even if an AT does not visually display, for instance, \textsc{Risk Subjects}, \textsc{Threat Objects}, and \textsc{Vulnerabilities}, it should refer to these elements in metadata. To understand this concretely, consider a particular implementation of this idea using Semantic Web technologies: an RDF database populating an OWL implementation of COVER (based on UFO OWL--- gUFO~\cite{almeida2019gufo}). Suppose ATs are conceptually defined according to the COVER theory of risk. In that case, constructing an AT out of this data involves extracting the necessary information through a SPARQL query--- for example, to pick certain \textsc{Threat Events} and \textsc{Qualities}. Capturing the AT elements in OWL is technically possible because OWL is strictly more expressive than the AT language (in terms of the mathematical structures it can represent). Accordingly, running usual AT algorithms for computing minimal attacks and security metrics becomes a matter of mapping the result of this query to an AT implementation language (for example, Galileo notation\footnote{The Galileo textual syntax was originally used for Fault Trees, but now an extended version of it is adopted to encode ATs. See: \url{https://dftbenchmarks.utwente.nl/at/at-format.html}.}) to generate an ordinary AT. The metadata guarantees that we can trace each AT element back to the well-founded knowledge graph, which explains \textit{why} a specific AT is the case. Finally, instead of the AT language, we propose that a graphical DSL designed according to those reference domain ontologies~\cite{oliveira2025toward} should be used for the conceptual modeling task of describing security risk scenarios. The resulting models should correspond to an RDF graph populating COVER OWL.

\paragraph{Summary.} Our proposed systematic approach~\cite{oliveira2025toward} involves a well-founded DSL for risk management, to be used by security analysts for conceptual modeling, storing data in RDF, and applying different techniques as a service of the DSL, including AT analyses, fault trees analysis, Bayesian analysis, etc. The outputs of these analyses provide different perspectives and can be input to other systems (e.g., security recommender systems). The AT formalism is to be used as a service of an adequate DSL (isomorphic to the domain ontology), not as a primary means for security modeling. One advantage of this approach resides in its \textit{modularity}: by providing a sound ontological ground for interoperability, we will be able to adopt different widespread techniques while, at the same time, enhancing their reasoning capabilities.

However, it is important to highlight that our analysis of ATs, based on UFO and COVER, does not rely on OWL or any Semantic Web technology. We take reference ontologies as conceptual artifacts/models that can provide clarity and greater disambiguation to ATs. These artifacts and their relationships to AT can be implemented using OWL, Prolog, Python, etc. For example, for the sake of argument, it is more essential to understand that AT nodes represent event types within a theory of risk than that, incidentally, event types are classes of an OWL ontology. Finally, our ontological analysis of AT has broader implications because similar ontological deficiencies hold for other risk management techniques, such as Fault Trees, Bayesian Networks, etc.


\section{Related work}
\label{section_related_work}


Several contributions address some of the points highlighted by us. We are closer to two categories: works that extend the standard AT formalism to account for its lack of expressivity (see Section \ref{subsection_semantic_incompleteness}); and works that offer AT modeling guidance (see Section \ref{subsection_limited_modeling_guidance}). Some of the most prominent are: dynamic ATs, in which the order of the BASes is expressed through sequential AND-gates \cite{jhawar2015attack,KRS15}; and attack-defense trees, which incorporate countermeasures/defenses \cite{kordy2011foundations}.

\paragraph{Popular attack tree extensions.} \textit{Dynamic Attack Trees} (DATs)~\cite{jhawar2015attack} extend static ATs by introducing a Sequential AND-gate, or SAND: for a SAND-gate to propagate the attack through the AT, all of its children elements must activate, but they need to do so in a left-to-right order. This gate introduces a notion of sequentiality--- or progress of time--- in the model, implying that some attack steps/goals have to happen/be reached before others for the attack to succeed. This requires a modification in the semantics of DATs in comparison to static ATs, where attacks need to be defined, e.g., as partially-ordered sets \cite{lopuhaa2022efficient}. However, as hinted in Section \ref{section_discussion_solution}, published works on DATs do not solve the underlying semantic confusion that we highlight with our ontological analysis, despite partially addressing the lack of expressivity in AT by introducing a new gate type--- which, however, increases its ambiguity. \textit{Attack-defense trees}~\cite{kordy2014attack} extend ATs by introducing a layer of defense nodes for the elements of the original AT that, if active, prevent the attacker from compromising the node to which they are linked. These defense nodes can be attacked by introducing new standard AT nodes that aim to compromise them. Although very useful in introducing and explicitly addressing defense mechanisms, attack-defense trees suffer from the same type of semantic overload highlighted in Section \ref{subsection_semantic_overload}, particularly concerning these newly introduced defense nodes.

\paragraph{AT modeling guidance.} The work in \cite{audinot2018guided} presents a framework to support a security expert in designing a pertinent AT for a given system. To do so, the authors start with an AT that is not fully refined and an explicit model of the real system to be analyzed, formalized as a transition system. While useful and promising in case this initial transition system exists, this method does not provide either general semantical guidelines to create an AT from scratch or an ontologically well-founded clarification of AT elements, which would help practitioners be more aware of the issues we highlight in Section \ref{section_contribution_ontological_analysis}. Authors in \cite{audinot2017my} develop a formal setting to assist experts in the design of ATs. As for \cite{audinot2018guided}, the system is described by an underlying model--- a finite state-transition system--- that reflects its dynamics and whose finite paths denote attack scenarios. While formalizing the labels of the AT nodes allows targeting the problem of imprecise or misleading text-based node names, this framework still relies on the presence of a transition system whose semantics w.r.t. the underlying system are assumed to be unambiguous and correct. As such, the ambiguity problems are only shifted from one model of the system to another without an ontologically well-founded clarificatory analysis.

Finally, the authors of \cite{pinchinat2020library} focus on the automatic synthesis of ATs by claiming that the design of ATs can be a tedious and error-prone process if done manually. While automation can be a powerful tool to minimize design errors in AT modeling, the approach adopted here is similar to the one of \cite{audinot2018guided}. It retains all the aforementioned strengths and limitations that distinguish the focus of their methodology from our work. 


\section{Final considerations}
\label{section_conclusion}


Attack trees have been used to describe security risk scenarios and calculate successful minimal attacks, and security metrics. Despite its immediate utility and being an easy-to-use domain-specific modeling language, the AT language displays numerous limitations due to its lack of ontological foundations. Through an ontological analysis grounded in the \textit{Common Ontology of Value and Risk} (COVER), we have shown issues related to semantic overload, ontological incompleteness, limited modeling guidance, and semantic interoperability affecting ATs. Our solution is a general systematic approach to risk management modeling, which bridges different communities: Formal Methods, Applied Ontology, Conceptual Modeling, Semantic Web, and Risk Management. Our research reinforces the idea that ``There’s no sense in being precise when you don't even know what you’re talking about.''\footnote{J. von Neumann, quoted by professional gambler B. Greenstein in his autobiography ``Ace on the River''.}. Modern real-world applications require more than formal rigor, they require ontological rigor for conceptual clarification and interoperability. 


\bibliographystyle{vancouver}
\bibliography{1-references}


\end{document}